\documentstyle[aps,prb,epsf]{revtex}

\begin{document}
\twocolumn[\hsize\textwidth\columnwidth\hsize\csname
@twocolumnfalse\endcsname

\title{Fractional vortices in the XY model with $\pi$ bonds}
\author{R. V. Kulkarni\cite{email}$^{1,2}$, E. Almaas$^1$, K. D. Fisher$^1$,
and D. Stroud$^1$}

\address{
$^1$Department of Physics, The Ohio State University,\\
Columbus, Ohio 43210 \\ $^2$Department of Physics, University of
California at Davis, \\Davis, California 95616
}

\date{\today}

\maketitle

\begin{abstract}
We define a new set of excitations in the XY model which we call
``fractional vortices''. In the frustrated XY model containing
$\pi$ bonds, we make the ansatz that the ground state configurations
can be characterized by pairs of oppositely charged fractional
vortices. For a chain of $\pi$ bonds, the ground state energy and the
phase configurations calculated on the basis of this ansatz agree well
with the results from direct numerical simulations. Finally, we
discuss the possible connection of these results to some recent
experiments by Kirtley {\it et al} \protect{[Phys. Rev. B {\bf 51},
R12057 (1995)]} on high-T$_c$ superconductors where fractional flux
trapping was observed along certain grain boundaries.
\end{abstract}

\draft 


\vskip1.5pc]

\section{Introduction}
The classical XY Hamiltonian is one of the most studied models in
statistical physics. In its usual, unfrustrated form, it is written
\begin{equation}
H = \sum_{\langle i j \rangle} J_{ij}[1 -\cos(\phi_i - \phi_j)],
\label{eq:xyham}
\end{equation}
where $\phi_i$ is a phase variable on the $i^{th}$ site ($0 \leq \phi<
2\pi$), the sum runs over distinct pairs $\langle ij \rangle$, and
$J_{ij}$ is the energy of the coupling between sites $i$ and $j$.  In
the ferromagnetic, nearest-neighbor case, $J_{ij}$ vanishes except
between nearest-neighbor sites and all the $J_{ij}$'s are equal to a
single positive constant $J$.  In this case, for spatial
dimensionality $d \geq 3$, there is a phase transition to a
ferromagnetic state at a critical temperature, with conventional
critical phenomena.  If $d = 2$, there is instead the
Kosterlitz-Thouless-Berezinskii phase transition, in which pairs of
oppositely charged integer vortices unbind at a finite temperature
$T_{KTB}$ \cite{kosterlitz}. The classical XY model has been found to
describe a wide variety of systems with complex scalar order
parameters, including bulk superconductors in $d = 3$, superconducting
films, Josephson junction arrays in $d = 2$, and superfluid He$_4$
films \cite{nelson}.

Recently, the XY model with {\em anti ferromagnetic} bonds i.e, with
bond strengths $J_{ij} < 0$ (also called {\em $\pi$ bonds}), has
received much attention
\cite{vannimenus,parker,gawiec,sigrist2,sigrist1,dominguez,bailey}, in
particular due to its possible relevance to high-T$_c$ superconductors
and other experimental systems \cite{sieger,katsumata,higgins}.
Specifically, if we consider the grain boundary between two high-T$_c$
superconductors with suitable misorientation of the crystalline axes,
then the resulting Josephson coupling across the boundary can have the
coupling energy $J_{ij}< 0$\cite{sigrist1}. This is a consequence of
the $d_{x^2-y^2}$ symmetry of the order parameter in many high-T$_c$
materials. Such grain-boundary interfaces have lately been studied in
a variety of experiments and in several geometries.  These experiments
have led to interesting results, such as the observation of the
trapping of half-integer and also other fractional flux quanta
\cite{kirtley1,kirtley2,tsuei}. These results can be explained using
models involving $\pi$ bonds
\cite{bailey,millis,geshkenbein,sigrist3,mints,kirtley3}.  Similar
models involving $\pi$ bonds have also been developed to explain such
phenomena as the paramagnetic Meissner effect \cite{dominguez}, also
observed in samples of high-T$_c$ superconductors.

A key concept in understanding the effects of $\pi$ bonds is
``frustration.''  Consider, for example, the XY model on a square
lattice with only the nearest-neighbor couplings nonvanishing.  If a
plaquette has an odd number of bonds, that plaquette is frustrated, in
the sense that no choice of angles in the four grains making up the
plaquette can simultaneously minimize {\em all} the bond energies.
Thus, a single $\pi$ bond will cause the two plaquettes adjoining that
$\pi$ bond to become frustrated.  In a {\em line} of $\pi$ bonds, only
the two plaquettes at the end of the line will become frustrated.
Because of the frustrated plaquettes, it is non-trivial to find the
ground state of the $XY$-model with $\pi$ bonds.  In this paper, we
will show, both numerically and by analytical arguments, that these
ground states are characterized by certain spatial phase configurations
which we call fractional vortices. We will also derive an
expression for the interaction energy of two fractional vortices in the
XY model.

The rest of the paper is organized as follows.  In Section II, we
define the fractional vortices and calculate the interaction energy of
a bound pair of fractional vortices for the XY model. In Section III,
we study the ground state of XY-lattices containing a single
$\pi$ bond, two $\pi$ bonds, and a string of $\pi$ bonds.  In each
case, using a variational ansatz for the ground-state configuration,
we find that there is a critical $\pi$ bond strength, above which the
ground state contains pairs of oppositely charged fractional vortices.
To check these results, we directly calculate the ground-state energy
of these lattices using a numerical relaxation technique based on the
equations of motion for overdamped Josephson junctions.  We find that
both the ground-state energy and the critical $\pi$ bond strength,
predicted by the variational approach, are in excellent agreement with
the numerical results.  Finally, in Section IV, we discuss the
possible relevance of these numerical results to experiments carried
out in systems containing $\pi$ junctions, such as high-T$_c$
superconductors containing grain boundaries and tricrystals, as
recently studied by Kirtley {\it et al}
\cite{kirtley1,kirtley2,tsuei}.
 
\section{Fractional Vortices in the Unfrustrated XY model}

Consider the Hamiltonian (\ref{eq:xyham}) for an XY model defined on a
square lattice with $N \times N$ sites.  If all the nearest-neighbor
couplings are equal, this may be written
\begin{equation}
H = J  \sum_{\langle ij\rangle}[1- \cos(\phi_i - \phi_j)]. 
\label{eq:xyham1}
\end{equation}
Hereafter, we shall use units such that $J = 1$.  The phase
angle, $\phi_{i}$, at point $(x_{i},y_{i})$ due to a fractional vortex
of charge $q$ at point $(x_{0},y_{0})$ is {\em defined} to be
\begin{equation}
\phi_{i}(x_0, y_0, q) = q \times \tan^{-1}(\frac{y_i - y_0}{x_i - x_0}).
\label{eq:fracvor}
\end{equation} 
For $q=1$, we recover the standard configuration for an integer vortex.
This definition can be seen as a generalization of the concept of
half-vortices introduced by Villain \cite{villain} for the same model.
Note that, while for the integer vortex the bond angles change
continuously, the fractional vortex case is characterized by a branch
cut, across which the bond angles are discontinuous.

This singularity leads to several other distinctions between integer
and fractional vortices.  For example, the energy associated with a
single integer vortex is proportional to $\ln(N)$. In the
thermodynamic limit, this is a weak divergence which makes the KTB
vortex-antivortex unbinding transition possible. By contrast, the
energy of an unbound fractional vortex is $\propto N$, since the
number of bonds along the branch cut is $\propto N$. Thus, it is
energetically unfavorable at all temperatures to create isolated
fractional vortices. But a bound pair of fractional vortices with
charges $q$ and $-q$ is much less expensive energetically, because
then the branch cut is restricted to the line joining the two charges:
the total energy should be proportional to the separation of the
fractional vortices. For fixed $q$ and large enough separations, this
energy is always larger than that of a pair of oppositely charged {\em
integer} vortices, whose energy varies as the logarithm of their
separation.  Nevertheless, for fixed separation, it is always possible
to find a non-integer $q$ such that that the energy of the fractional
vortex pair is less than the corresponding energy for the integer
vortices. In the following, we derive expressions for the energy of a
bound pair of fractional vortices in the XY model, and compare them to
numerical results obtained by calculating the energy explicitly for
these configurations.

\begin{figure}[tb]
\epsfysize=8cm
\centerline{\epsffile{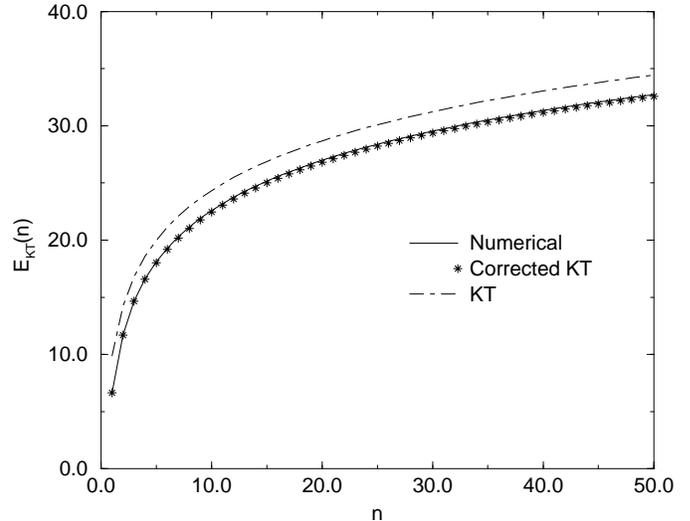}}
\caption{Calculated energy of a bound pair of integer vortices as a
      function of separation $n$, in units of $J$.  Full curve:
      numerically exact results.  Dot-dashed curve: KT-approximation
      [Eq.\ (\ref{eq:KT})].  Asterisks: KT-approximation plus core
      correction.}
\label{fig1}
\end{figure}
We first consider a bound pair of integer vortices of charge $\pm 1$
located at ($x_0$, $y_0$) and ($x_1$, $y_1$).  The standard 
KT expression for the energy of the pair is obtained by approximating
the Hamiltonian as
\begin{equation}
H \sim  \frac{1}{2} \sum_{<ij>}  (\phi_{i} - \phi_{j})^{2}. 
\label{eq:happrox}
\end{equation}
For the phase configuration, we use $\phi_i = \phi_i(x_0, y_0, +1) +
\phi_i(x_1, y_1, -1)$, where $x_1-x_0 = n$ and $y_1-y_0 = 0$ (in units
of the lattice constant $a$).  Substituting this configuration into
(\ref{eq:happrox}) gives the Kosterlitz-Thouless formula for the
interaction energy, $E_{KT}$, of two oppositely charged integer
vortices:
\begin{equation}
E_{KT}(n) = 2 \pi \left[ \ln n + \frac{\pi}{2}\right].      \label{eq:KT}
\end{equation}
In Fig.\ \ref{fig1}, we compare this expression to the energy of a
pair of oppositely charged integer vortices, computed using the same
phase configuration but the exact $H$.  The discrepancy arises from
the expansion of the cosine factor, which is inaccurate for the bonds
closest to the vortices.  This inaccuracy is remedied by a making a
core correction i.e, by calculating the contribution from the bonds on
the perimeter of the plaquettes surrounding the vortices exactly,
rather than by a quadratic expansion.  For large $n$, the
core-correction energy, $E_c(n)$, is approximately given by
\begin{equation}
E_c(n) = \pi^2 - 8 + \frac{12- \pi^2}{2n^2} +
\frac{8+\pi^2}{16n^4}.
\end{equation}
As can be seen from Fig.\ \ref{fig1}, the numerically calculated
energy is well approximated by $E_{KT}(n) + E_c(n)$.
\begin{figure}[tb]
\epsfysize=6cm
\centerline{\epsffile{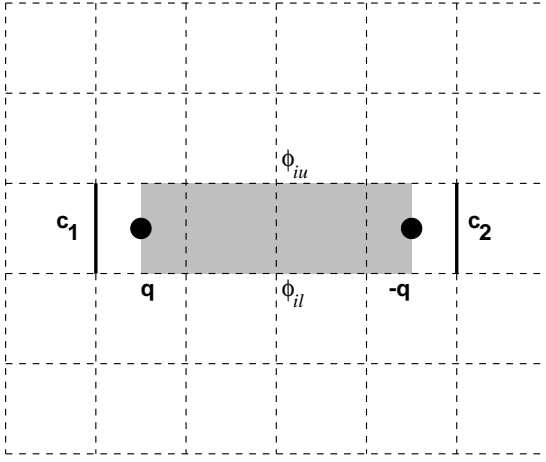}}
\vspace*{.2cm}
\caption{Schematic drawing of a bound pair of fractional vortices
      arranged parallel to the $x$-axis.  For this configuration, the
      vortices are separated by three plaquettes.  $\phi_{iu}$ and
      $\phi_{il}$ are the phases at the two ends of bonds in the
      shaded region (region A); the angle $\theta_{i,A}$ in the text
      is defined by $\theta_{i,A} \equiv \phi_{iu} - \phi_{il}$.  Core
      corrections are calculated only for the bonds denoted $C_1$ and
      $C_2$, as discussed in the text.}
\label{fig2}
\end{figure}
Next, we calculate the energy $E(q, n)$ of a pair of fractional
vortices $\pm q$, separated by a distance $n$ in the $x$-direction, as
shown schematically in Fig.\ \ref{fig2}.  Note that $E(1,n)$ is simply
the interaction energy of a pair of integer vortices, as just
discussed.  The phase configuration of this pair is then given by
$\phi_i = \phi_i(x_0, y_0, q) + \phi_i(x_1, y_1, -q)$, where $x_1-x_0
= n$ and $y_1-y_0 = 0$.  We now divide the bonds into two groups: (A)
those intersected by the line segment joining the two vortex centers;
and (B) the remaining bonds.  Let $E_{A}(q,n)$ and $E_{B}(q,n)$ be the
corresponding energy contributions to $E(q, n)$ coming from these two
groups of bonds.  Thus $E(q, n) = E_A(q, n) + E_B(q, n)$.  To obtain
$E(q, n)$ we proceed as follows:

(1). We calculate $E_{A}(1,n)$, using the quadratic expansion for the
cosine. $E_A(1,n)$ is simply the contribution to the total energy of a
bound pair of integer vortices arising from the bonds along the branch
cut. Note that there are $n$ bonds in region A. Once $E_A(1,n)$ is
known, we get $E_{B}(1,n) = E_{KT}(n) - E_{A}(1,n)$.

(2). We obtain $E_{B}(q,n)$ by noting that, in the quadratic
approximation, $ E_{B}(q,n) = q^2 E_{B}(1,n)$.

(3). Finally, we determine $E_{A}(q,n)$ by directly evaluating it
using the full expression for the cosine, not the quadratic expansion.
This is necessary, because the bond angles in region A are not small
for arbitrary $q$.

We now use the outlined procedure to obtain $E(q,n)$.

{\em Step 1}: Let $\theta_{i,A}(1,n) \equiv \phi_{iu} - \phi_{il}$
denote the $i^{th}$ {\em bond} angle (cf. Fig.\ \ref{fig2}) in region
A for q = 1.  For the two-vortex configuration, $\theta_{i,A}(1, n)$
is given by
\begin{equation}
\theta_{i,A}(1,n)\! =\! 2\left[ \tan^{-1}\!\! \left( \frac{1}{2i\!-\!1}
  \right)\! + \! \tan^{-1}\!\! \left(\frac{1}{2n\!  -\!2i\!  +1} \right)
  \right]
\end{equation}
Using the quadratic approximation, the corresponding energy
contribution $E_{A}(1,n)$ is given by
\begin{equation}
 E_{A}(1,n) = \frac{1}{2} \sum_{i=1}^{n} \theta_{i,A}^2(1,n).
 \label{eq:Ea1}
\end{equation}
Using the approximation: $~\tan^{-1}\left[1/(2i-1)\right] \sim
1/(2i-1) $ for $ i \ge 2 $, we find
\begin{eqnarray}
E_{A}(1,n)\! &=&\! \frac{\pi^2}{2}\! +\! \left(\frac{\pi}{2}\! 
+\! \frac{2}{2n -1} \right)^2 \!+\! \frac{2}{n}\left[\gamma \!+\! 2\ln 2 \!
-\! 2\! -\! 2n\right] \nonumber \\
 &+& \frac{2}{n}\psi\left(n - \frac{1}{2}\right)
 - \psi^{\prime}\left(n - \frac{1}{2}\right) \nonumber   \\
E_{B}(1,n) &=& E_{KT}(n)  - E_{A}(1,n), 
\end{eqnarray}
where $\psi(x)$ and $\psi^\prime(x)$ are the Digamma function and its
derivative, $\gamma$ is Euler's constant, and $E_{KT}$(n) is given by
Eq.\ (\ref{eq:KT}).
 
{\em Step 2.}  Using the results of step 1 and Eq.\ (\ref{eq:KT}), we
get
\begin{equation}
E_{B}(q,n) = q^{2} \left[2 \pi \ln n + \pi^2 - E_{A}(1,n)\right].
\label{eq:Eb}
\end{equation}

{\em Step 3.}  The next step is to calculate $E_A(q, n)$. Bond angles
in region $A$ are given by
\begin{equation}
\theta_{i,A}(q,n) =  q \left[2 \pi - \theta_{i,A}(1,n)\right].
\end{equation}
Correspondingly, the energy $E_{A}(q,n)$ is given by 
\begin{equation}
E_{A}(q,n) =  \sum_{i=1}^{n} (1 -\cos\left[\theta_{i,A}(q,n)\right]).
\label{eq:Ea}
\end{equation}
Since the bond angle $\theta_{i,A}(q,n)$ is not small for an arbitrary
$q$, we cannot expand the cosine term only to second order.  But for
{\em any} $q$, the difference $\theta_{i,A}(q,n) - \theta_{n/2,A}(q,n)$
{\em is} a small parameter for any $i \ge 2$.  Expanding the cosine
term in Eq.\ (\ref{eq:Ea}) to second order in this parameter, we
obtain an expression for $E_{A}(q,n)$.  This expression can be summed,
and eventually gives

\[
E_{A}(q,n) = (n-2)\left[(1 - \cos\alpha_n ) \!+\! \frac{8q^2}{n^2}
	\cos\alpha_n \!+\! \frac{4q}{n} \sin\alpha_n \right] \]
\[+2\left\{ 1\! -\! \cos \left[ q\left( \frac{3\pi}{2} \!-\! \frac{2}{2n\!-\!1}
	\right) \right] \right\} \!-\! \left[\frac{4q^2}{n} \cos\alpha_n + 
	q \sin\alpha_n\right] \]
\begin{equation}
\times\sum_{m=2}^{n-1} \theta_{m,A}(1,n) +\frac{q^2}{2} \cos\alpha_n 
	\sum_{m=2}^{n-1} \theta_{m,A}^2(1,n),
\end{equation}
where
\begin{eqnarray}
\sum_{m=2}^{n-1}\theta_{m,A}(1,n)=&& 2\gamma - 4 + 4 \ln 2 + 2 \psi \left(n-
	\frac{1}{2}\right), \\
\sum_{m=2}^{n-1} \theta_{m,A}^2(1,n)=&& \pi^2 +\frac{4}{n} \left( \gamma + 
2 \ln 2  - 2 - 2n \right) \nonumber \\
+ \frac{4}{n} \psi && \left( n - \frac{1}{2} \right) - 2 \psi^{\prime} \left(n 
- \frac{1}{2} \right) \\
\alpha_n = q&& \left(2\pi - \frac{4}{n}\right).
\end{eqnarray}
\newpage

\widetext
\begin{figure}[tb]
\epsfysize=7.9cm
\centerline{\epsffile{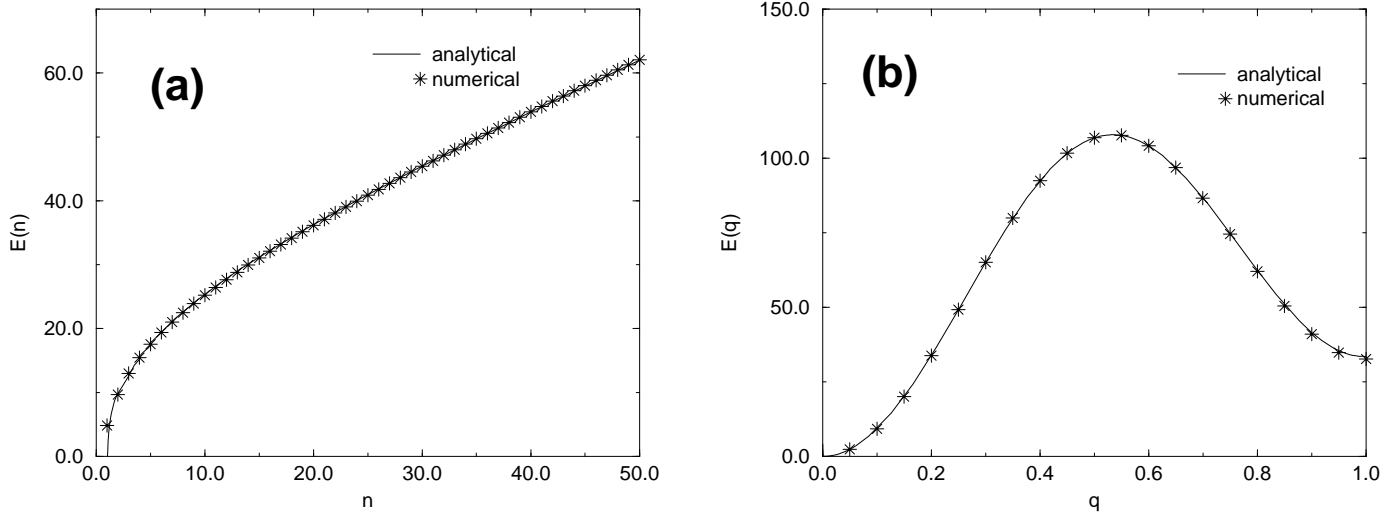}}
\caption{Energy of a bound pair of fractional vortices in an
      unfrustrated XY-lattice as obtained numerically ($\ast$) and
      from the analytical approximation, Eq.\ (\ref{eq:eqn1}) (full
      line), for (a) fixed charge, $q = 0.8$, as a function of
      separation; and (b) fixed separation, $n=50$, as a function of
      charge.  }
\label{fig3}
\end{figure}
\narrowtext

Adding up the contribution from the two regions, we finally get the
required expression for the energy of a bound pair of fractional
vortices.  As noted earlier, the core corrections must be included to
attain high numerical accuracy. In the present case, it is sufficient
to include these corrections only for the bonds labeled C$_1$ and
C$_2$ in Fig.\ \ref{fig2}, using the procedure outlined earlier. This
approximation is equivalent to extending region A to include bonds
C$_1$ and C$_2$ . Correspondingly, the core-corrected total energy is
given by
\begin{equation}
E(q,n) \!=\!  E_{A}(q,n) \!+\! E_{B}(q,n)\! -\! q^{2} \theta_c^{2}
         \!+\!  2 \left[1 \!-\! \cos(q \theta_c )\right]
\label{eq:eqn1}
\end{equation}
where 
\begin{equation}
\theta_c ~=~ \frac{\pi}{2} - \frac{2}{2n +1}.
\label{eq:eqn2}
\end{equation}
 
Expressions (\ref{eq:eqn1}) and (\ref{eq:eqn2}) are compared to the
results of numerical computation in Figs.\ \ref{fig3}(a) and
\ref{fig3}(b); agreement between the two is excellent.  On the basis
of this agreement, which is equally good for all values of $q$ and $n$
which we have considered, we present this result as a good analytical
expression for the interaction energy between two fractional vortices
in the unfrustrated XY model on a square lattice.  This result is a
generalization of the integer vortex excitations proposed by
Kosterlitz and Thouless.

For large $n$, we can further simplify the above expression by
dropping terms of ${\cal O}(1/n)$ and smaller in Eq.\ (\ref{eq:eqn1})
to get
\begin{eqnarray}
E(q,n) &=& (n-2)[1\! -\! \cos(2 \pi q)] + 2\ln n
\left[\pi q^2 \!-\! q \sin(2\pi q)\right]
\nonumber \\
    &&+~ \frac{3}{4} \pi^2 q^2 ~+~ 2\left[1 - \cos(\frac{3 \pi q}{2})\right] .
\end{eqnarray}

\vspace*{8.7cm}

\section{Fractional Vortices in the XY model With $\pi$ bonds} 
 
The fractional vortex configurations introduced in the previous
section provide a natural way of characterizing the ground state of
the XY model containing $\pi$ bonds.  In this section, we implement
this description by making a variational guess for the ground-state
configuration using fractional vortices.  We then compare our
variational results with those obtained by numerically relaxing to the
ground-state configuration, and find excellent agreement.  In the
following subsections, we will focus on obtaining the critical
bond strength, $\lambda_c$, above which the ferromagnetic ground-state
solution becomes unstable, and the ground-state configuration contains
bound pairs of fractional vortices. For the case of one and two
$\pi$ bonds, we also compare our results to those from previous
studies by Vannimenus {\it et al} \cite{vannimenus}. Note that in
these calculations, in which the goal is to calculate the threshold
bond strength above which the ferromagnetic ground state becomes
unstable rather than the absolute energies as a function of $\lambda$,
it is unnecessary to include the core corrections. Hence, $\lambda_c$
can be calculated analytically as demonstrated below.  The
ground-state configuration and energy for $\lambda > \lambda_c$ do
need the core corrections for greatest accuracy. We obtain them
numerically using our variational guess and discuss them in the
subsequent section.

\subsection{One $\pi$ bond}

We first consider the case of a single $\pi$ bond i.e, a single
antiferromagnetic bond in a host of ferromagnetic bonds. 

\newpage

\widetext

\begin{figure}[tb]
\epsfysize=8cm
\centerline{\epsffile{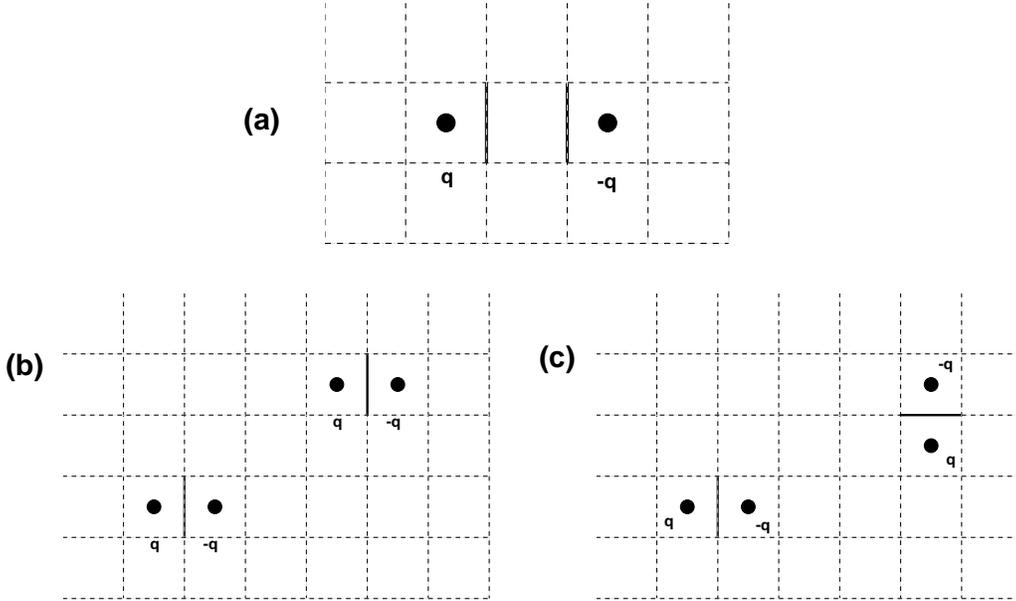}}
\vspace*{.1cm}
\caption{Schematic plot of three configurations, each containing
      $\pi$ bonds (full lines), which are (a) parallel and adjacent,
      (b) parallel and non-adjacent, and (c) perpendicular and
      non-adjacent.  Also shown are the corresponding locations of the
      fractional vortex charges for the variational configurations.  }
\label{fig4}
\end{figure}

\narrowtext

\noindent
As before, we take the bond strength of the ferromagnetic bonds to
equal unity, and we denoted the strength of the antiferromagnetic bond
by $\lambda$ ($\lambda \ge 0$).  The problem is to obtain the
ground-state configuration and energy for arbitrary strength,
$\lambda$, of the $\pi$ bond.

To solve this problem, we make a variational guess for the
ground-state configuration: it is the phase configuration
corresponding to a bound pair of fractional vortices of strength $\pm
q$, located at the centers of the two plaquettes adjacent to the
$\pi$ bond.  The charge, $q$, is a variational parameter with respect
to which the ground-state energy is minimized for a given $\lambda$.

The total energy of this configuration discussed above is readily
obtained using the procedure of the previous section, suitably
corrected for the fact that we have a $\pi$ bond instead of a
ferromagnetic bond.  The angle difference across the $\pi$ bond is
given by $ \theta_{\pi} = q \pi $.  Then, using Eqs.\ (\ref{eq:Ea1})
and (\ref{eq:Eb}), we get
\begin{equation}
E_{B}(q) = \frac{1}{2} q^{2} \pi^{2} ,
\end{equation}
while from Eq.\ (\ref{eq:Ea}) we find
\begin{equation}
E_{A}(q) = 1 + \lambda \cos(q \pi) .
\end{equation}
Adding these two terms gives the total energy of the configuration.
Minimizing this energy with respect to $q$ yields the condition
\begin{equation}
q \pi = \lambda~ \sin(q \pi).
\label{eq:onebond}
\end{equation}

For $\lambda \le 1 $, the ground-state configuration corresponds to $q
= 0$: all the phases are perfectly aligned.  For $\lambda > 1$, 
\vspace*{9.2cm}

\noindent
the ground-state configuration corresponds to a bound pair of
fractional vortices with charges $\pm q$ obtained by solving Eq.\
(\ref{eq:onebond}).  Thus, the ferromagnetic ground state is unstable
above a critical bond-strength value $\lambda_c = 1$.  The same value
has been obtained previously by workers using different approaches
\cite{vannimenus,parker}.

\subsection{Two $\pi$ bonds }

We now consider the case of two parallel, adjacent $\pi$ bonds.  As
before, our variational guess for the ground state is the
configuration corresponding to a bound pair of fractional vortices; we
take these to be located as shown in Fig.\ \ref{fig4}(a).  The
corresponding total energy is again calculated using the procedure
outlined in Section II. Using Eqs.\ (\ref{eq:Ea1}) and (\ref{eq:Eb}),
the energy contribution $E_B(q)$ is
\begin{equation}
E_{B}(q) \!=\! q^{2} \left\{ 2\pi \ln 2 \!+\! \pi^2 \! -\! 
\left[\frac{\pi}{2}\! +\! 2\tan^{-1}(1/3)\right]^2\right\}.
\end{equation}
Using Eq.\ (\ref{eq:Ea}), we get  
\begin{equation}
E_{A}(q)  = 2 + 2 \lambda \cos\left\{2q\left[3\pi/4
-\tan^{-1}(1/3)\right]\right\}.
\end{equation}
Adding these two contributions gives the total energy, which is to be
minimized with respect to $q$ for a given $\lambda$.  This procedure
gives the critical value $ \lambda_c = 0.563$, which is in good
agreement with the exact value $ \lambda_c = \pi/2 - 1$,
obtained by Vannimenus {\it et al}\cite{vannimenus}.

Next, we consider the case of two parallel, but non-adjacent
$\pi$ bonds, as shown in Fig.\ \ref{fig4}(b). Taking the bond centers
to have the coordinates $(0,0)$ and $(m,n)$, 

\begin{figure}[tb]
\epsfysize=4.5cm
\centerline{\epsffile{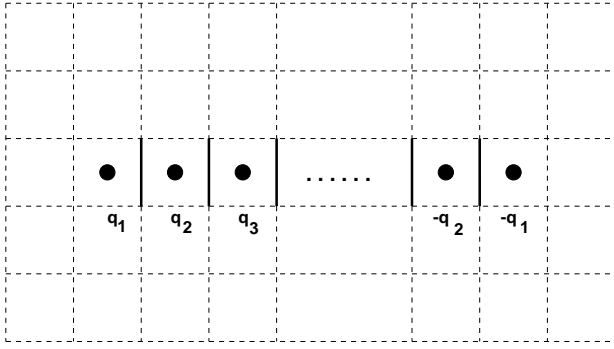}}
\vspace*{.1cm}
\caption{Schematic plot of the assumed variational configuration
      containing $m$ pairs of fractional charges, for a chain of
      $\pi$ bonds (solid line segments) of length $2m$.}
\label{fig5}
\end{figure}

\noindent
we calculate the energy using the variational procedure described
above with vortex charges as shown.  For large separation between the
bonds ($\sqrt{m^2 + n^2} \gg 1$), this procedure gives
\begin{equation}
E_B(q) = q^2 \left[2\pi^2 - 2\alpha_{mn} -  ( \pi - \alpha_{mn})^2\right]
\end{equation}
and 
\begin{equation} 
E_{A}(q) = 2 + 2\lambda \cos\left[q (\pi + \alpha_{mn})\right],
\end{equation}
where 
\begin{equation}
\alpha_{mn} = \frac{ m^2 - n^2 }{ (m^2 + n^2)^2 } .
\end{equation}
Minimizing the total energy gives the critical bond strength as
\begin{equation}
\lambda_c = \frac{ 1 - 2 \alpha_{mn}/{\pi}}{1 + 2 \alpha_{mn}/{\pi}}.
\end{equation} 

Similarly, for two non-adjacent {\em perpendicular} $\pi$ bonds [Fig.\
\ref{fig4}(c)], we find
\begin{equation}
E_{B}(q) = q^2 \left[ 2 \pi^2 - 2 \beta_{mn} - ( \pi - \beta_{mn} )^2\right]
\end{equation}
and 
\begin{equation}
E_{A}(q) = 2  + 2\lambda \cos\left[ q ( \pi + \beta_{mn} )\right] ,
\end{equation}
where 
\begin{equation} 
\beta_{mn} = \frac{2mn}{(m^2 + n^2 )^2 }.
\end{equation}
In this case, the critical bond strength is  
\begin{equation}
\lambda_c = \frac{ 1 - \beta_{mn}/\pi} { 1 + \beta_{mn}/\pi }.
\end{equation}

These results are identical to those obtained previously by Vannimenus
{\it et al}, using a different approach \cite{vannimenus}. The
agreement lends support to our hypothesis that the ground-state
configuration of such systems can be characterized by a set of
fractional vortices (in the cases considered here, a set of only two
oppositely charged fractional 

\begin{figure}[tb]
\epsfysize=8.3cm
\centerline{\epsffile{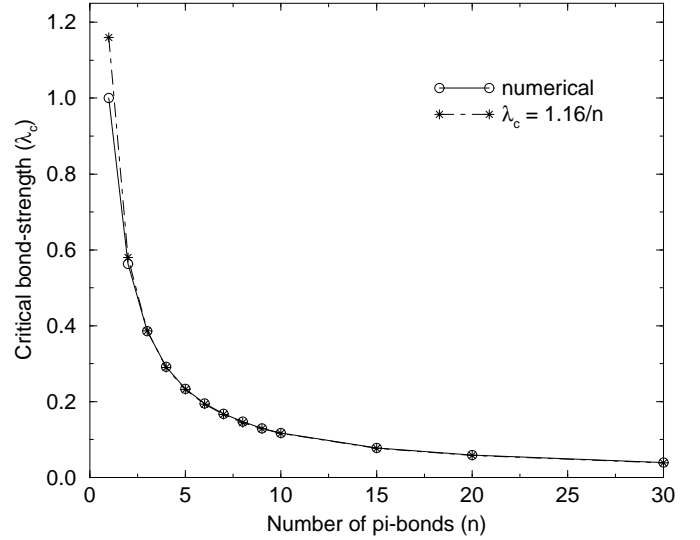}}
\caption{Critical bond strength $\lambda_c$ as function of chain
      length, for a chain of $\pi$ bonds of length $n$.  Open circles
      and full curve: numerical results.  Asterisks and dashed curve:
      analytical approximation, $ \lambda_c = \frac{1.16}{n}$.  }
\label{fig6}
\end{figure}

\noindent
vortices).  Besides having the merit of
simplicity, our approach also easily yields the ground-state
configuration and energy for arbitrary $\lambda$.  Moreover, our
procedure can be used to obtain the ground-state even when the
separation between the bonds is not large.  In particular, for two
parallel bonds such that $m=n=1$, our variational procedure yields the
surprising result that $\lambda_c =1$ for this configuration.  The
same result was obtained in a numerical study done by Gawiec {\it et
al}\cite{gawiec}.  Finally, our variational ansatz can readily be
generalized to longer $\pi$ bond chains, as we shall see in the next
section.

\subsection{ Chains of $\pi$ bonds }

Next, we consider chains of $\pi$ bonds of length $n \ge 3 $.  In this
case, we make the variational ansatz that the ground state consists of
$n/2$ or $(n+1)/2$ pairs of oppositely charged fractional vortices for
even or odd $n$, arranged as shown in Fig.\ \ref{fig5}.  As before, we
proceed by calculating the contribution to the total energy from
regions A and B.  However, the procedure outlined in Section II has to
be generalized to include many pairs of fractional vortices.  Since
the details are significantly different from that outlined in section
II, we briefly describe the generalized procedure below.

(1). We consider the case in which all charges have magnitude unity.
The total energy of this configuration is given simply by the KT
expression
\begin{equation}
E_{KT} = -2 \pi \sum_{j < k} q_j q_k \ln(n_{jk}) + \pi^{2}
          \sum_j q_j^2,
\end{equation}

\begin{figure}[tb]
\epsfysize=8.1cm
\centerline{\epsffile{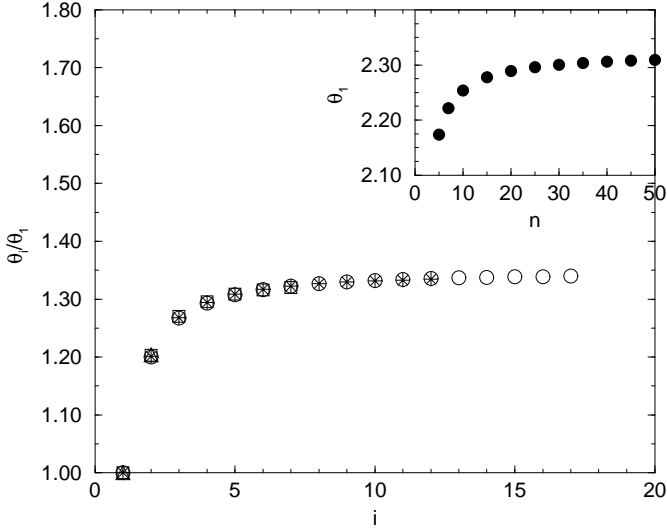}}
\caption{Ratio of $i^{th}$ bond angle, $\theta_i$, to the first
      bond angle $\theta_1$, plotted as a function of $i$ for
      chain lengths $n = 5 (\triangle), 15 (\Box), 25 (\ast), 35
      (\circ)$. The bond strength is $\lambda = 1$. Inset: variation
      of $\theta_1$ with chain length $n$, for $\lambda = 1$.}
\label{fig7}
\end{figure}

\noindent
where $n_{jk}$ is the distance between the charges $q_j$ and $q_k$,
and the second sum runs over all the individual charges, each of which
has magnitude unity. This result is obtained by using a small-angle
expansion for the contribution from each bond angle difference
$\theta_b$, and summing those contributions to give
\begin{equation}
E_{KT}  = \frac{1}{2} \sum_{b}(\theta_b)^2.
\end{equation}
The bond angle $\theta_b$ is, in turn, decomposed as
\begin{equation} 
\theta_b = \sum_{k} q_k \theta_{k,b}    \label{eq:ba},
\end{equation}
where $k$ labels the position of the charges and $\theta_{k,b}$ is the
contribution to $\theta_b $ from a charge of unit magnitude at $k$.

(2). Next, we consider the case in which the charges are fractional
($|q| < 1$).  The bonds can still be divided into classes A and B as
discussed earlier.  In the case of fractional charges, the bond angle
differences in region B are still given by Eq.\ (\ref{eq:ba}).
Correspondingly, the energy contribution from bonds in region B is
\begin{eqnarray}
E_B & = &\frac{1}{2} \sum_{b}^{B} \theta_{b}^2  \nonumber \\
    & =  &  \frac{1}{2} \sum_{b}^{A + B}\theta_{b}^2 - 
\sum_{b}^{A} \theta_{b}^2,
\end{eqnarray}
where $\sum_b^B$ and $\sum_b^A$ denote sums over all bonds in region B
and in region A.

\begin{figure}[tb]
\epsfysize=8.1cm
\centerline{\epsffile{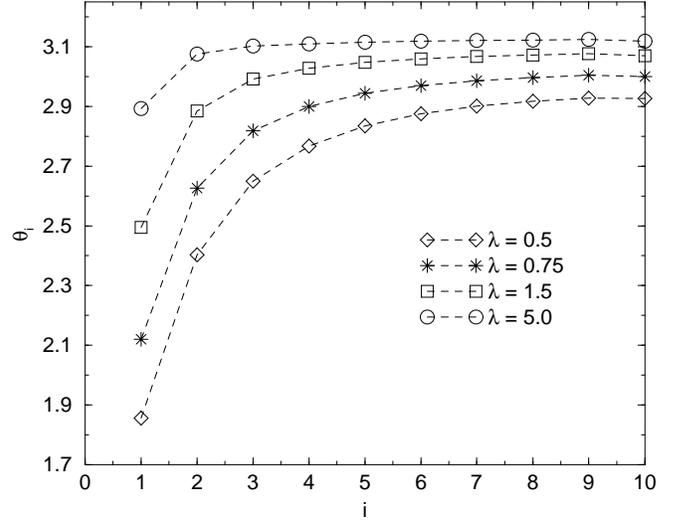}}
\caption{The $i^{th}$ bond angle, $\theta_i$, plotted as a function of
      $i$ for bond strengths $\lambda = 0.5, 1.0, 1.5, 2.0 $.  The
      chain length, $n$, is taken to be 20.}
\label{fig8}
\end{figure}

(3). Finally we calculate the energy $E_A$, using suitable expressions
for the bond angles in region A, as obtained from the multi-vortex
configuration but without making the small angle expansion.

To minimize the resulting total energy, which is a function of all the
$q_k$'s, we used two procedures: (i) Powell's multidimensional
direction set method\cite{powell}, and (ii) a genetic
algorithm\cite{genetic}.  Both methods successfully converged to the
same minimum energy and configuration, from which we deduced the
critical bond strength, $\lambda_c$, for various values of $n$. Fig.\
\ref{fig6} shows our results for $\lambda_c(n)$.  As can be seen from
the main part of the figure, it fits very well to the approximate
expression $\lambda_c \approx 1.16/n$.  A consequence of this $1/n$
dependence is: if the system has a finite concentration of $\pi$ bonds
randomly distributed in the lattice, then in the thermodynamic limit
$\lambda_c \rightarrow 0$. This behavior follows from the fact that,
in the thermodynamic limit, there is always a finite probability of
having an arbitrarily large chain length $n$, and hence an arbitrarily
small $\lambda_c$.

We now look at the variations in the bond angles along the $\pi$ bond
chain as a function of chain length, $n$, and bond strength, $\lambda$
(for $\lambda > \lambda_c$). First, we discuss the variation with
fixed bond strength, taking $\lambda = 1$.  Fig.\ \ref{fig7} shows the
ratio of the bond angles, $\theta_i$, along the chain (not including
the central bond) to the bond angle across the first $\pi$ bond,
$\theta_1$, as a function of position along the chain for various
chain lengths. A number of features deserve mention.  First, for a
chain of length $n = 2m$, the ratio of the bond angles,
$\theta_{i}/\theta_1$, for $ i < m$ is independent of $m$.  Second,
since the bond angle distribution is symmetric, we only need to look
at the bonds in the range $1 \le i \le m$.  Third, the bond angles
increase monotonically as one moves along the chain from its edges to

\begin{figure}[tb]
\epsfysize=8.1cm
\centerline{\epsffile{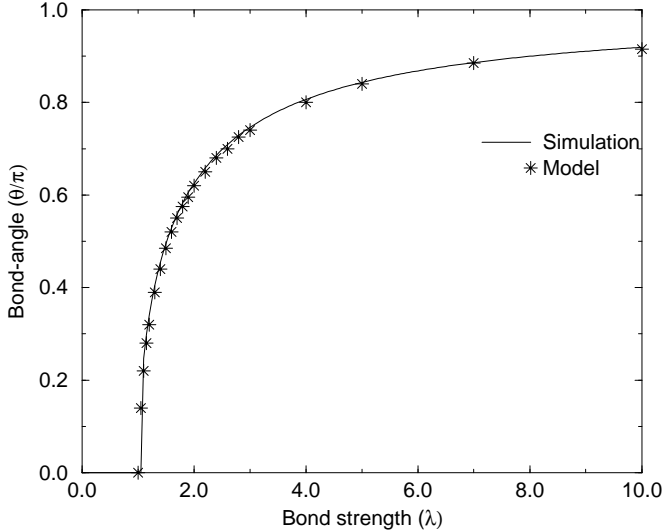}}
\caption{Bond angle, $\theta$, as function of bond strength $\lambda$
      for a single $\pi$ bond obtained using (a) fractional-charge
      variational ansatz ($\ast$), and (b) numerical simulations
      (solid line).}
\label{fig9}
\end{figure}

\noindent
the center i.e, as $i$ increases from 1 to $m$.  The inset shows the
variation of $\theta_1$ with chain length for $\lambda = 1$.  Note
that with this choice of $\lambda$, the bond angles saturate quickly:
they are roughly constant over the ``interior bonds,'' such that $i
\ge 3$.  Furthermore, this constant value (approximated by the central
bond angle) approaches $\pi$ as the chain length $n$ increases.

Fig.\ \ref{fig8} shows how the bond angles, $\theta_i$, vary with bond
strength, $\lambda$, for a fixed chain length ($n = 20$). As already
seen in the previous figure, $\theta_i$ rapidly tends to saturate
towards its central value with increasing $i$.  Moreover, the central
bond angle quickly increases from $0$ to $\pi$ as $\lambda$ increases
for $\lambda > \lambda_c$. Thus, we can `tune' the central bond angle
to any desired fraction of $\pi$ by appropriately adjusting $\lambda$.

Although the underlying variables are the $\theta_i$'s, it is of
interest to mention corresponding trends in the fractional
charges. For small $\lambda$, these charges decrease monotonically
with increasing $i$, so that the largest charges reside at the ends of
the chain. For larger $\lambda$, charges comparable in magnitude to
those at the ends appear away from the ends.
 
\subsection{Numerical Check of Variational Procedure.}  

To check our variational approach, we have carried out an independent
minimization to calculate the ground-state energy of the system
containing $\pi$ bonds \cite{fisher}, {\em without} making any
assumptions about the presence or absence of fractional vortices.  To
carry out this minimization, we imagine that the $ij^{th}$ bond
is actually an overdamped Josephson junction connecting nodes $i$ and
$j$.  The current flowing through that bond from node $i$ to node $j$

\begin{figure}[tb]
\epsfysize=8.1cm
\centerline{\epsffile{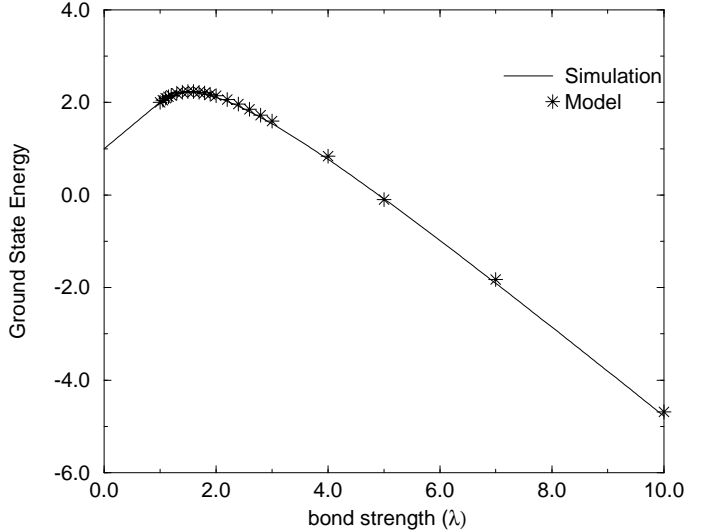}}
\caption{Total energy as function of $\lambda$ for a single $\pi$ bond
      obtained using (a) variational ansatz ($\ast$), and (b)
      numerical simulations (solid line).}
\label{fig10}
\end{figure}

\noindent
is then
\begin{equation}
I_{ij} = I_{c,ij} \sin{\left(\phi_i-\phi_j\right)} +
        \frac{\hbar}{2eR_{ij}}\frac{d}{dt}\left(\theta_i-\theta_j\right),
\label{eq:rsj1}
\end{equation}
and the sum of these currents must equal the total external current,
$I_i^{ext}$, fed into node $i$:
\begin{equation}
\sum_b{I_{ij}} = I_i^{ext}.
\label{eq:rsj2}
\end{equation}
where $I_{c,ij}$ is the critical current of the junction between
grains $i$ and $j$, and $R_{ij}$ is the corresponding shunt
resistance.  These equations can be put into dimensionless form using
the definitions $i_{ij} \equiv I_{ij}/I_c$ and $g_{ij} \equiv
R/R_{ij}$, where $I_c$ and $R$ are a convenient normalizing critical
current and shunt resistance, and introducing the natural time unit
$\tau \equiv \hbar/(2eRI_c)$.  Combining these equations yields a set
of coupled ordinary differential equations which is easily reduced to
matrix form and solved numerically, as described by many previous
investigators\cite{yu}.  For our work, we employed a fourth-fifth
order Runge-Kutta integration with variable time step.

For present purposes, we are interested, not in examining the
dynamical properties of arrays with $\pi$ bonds, but rather in finding
the minimum-energy configuration of such arrays.  To that end, we have
simply iterated this set of coupled equations of motion, with {\em no}
external current, allowing the phases to evolve until they reach a
time-independent configuration.  As has been shown by previous
workers, this configuration will correspond to a local minimum-energy
state of the corresponding Hamiltonian $H = -\sum_{\langle ij \rangle}
(\hbar I_{c;ij}/2e)\cos(\phi_i - \phi_j)$.  We then compare the
resulting configuration and energy with those predicted by the
fractional vortex variational ansatz for the ground state.

\begin{figure}[tb]
\epsfysize=8.1cm
\centerline{\epsffile{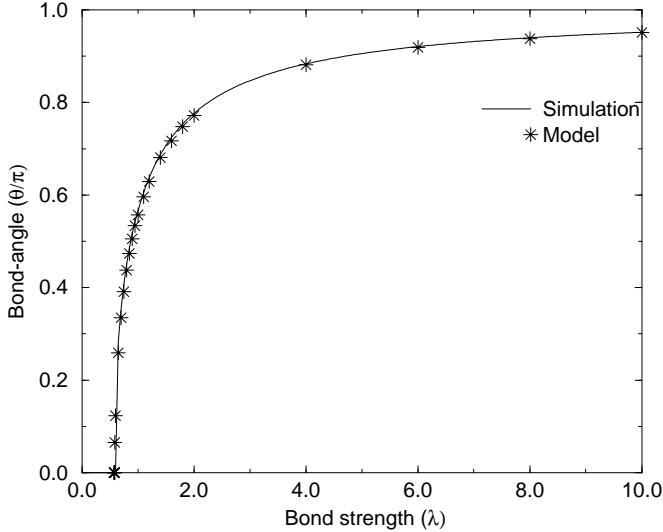}}
\caption{Same as Fig.\ \ref{fig9} but for two adjacent $\pi$ bonds.}
\label{fig11}
\end{figure}

To make the comparison as straightforward as possible, we made the
simplifying assumption $R_{ij} = R$ for all Josephson junctions,
whether $0$ or $\pi$. We took $i_{c,ij} = 1$ for all normal junctions,
and $i_{c;ij} = -\lambda$ for all $\pi$ junctions.  Since no external
current is to be applied to the system, we carry out these
calculations using square arrays of junctions with periodic boundary
conditions in both directions.

Each simulation begins with phases randomized at each grain.  The
system is then relaxed according to the Eqs.\ (\ref{eq:rsj1}) and
(\ref{eq:rsj2}) for an interval of $50-100\tau$.  We then evaluate the
final energy, as well as the phase difference, $\theta_{ij}$, across
each $\pi$ junction. Once equilibrium is reached for a given
$\lambda$, we increment or decrement $\lambda$, and the system is
allowed to relax again without re-randomizing the phases. Even quite
large arrays (50 $\times$ 50 plaquettes) relax quite quickly using
this procedure, except near the critical point, but care must be taken
to avoid taking data from simulations in which the system is trapped
in a metastable state. We have used arrays ranging from 10 $\times$ 10
plaquettes to 50 $\times$ 50, and occasionally as large as 70 $\times$
70 to examine convergence of equilibrium values.
   
Figs.\ \ref{fig9} and \ref{fig10} show the the exact ground-state
energy and corresponding $q$ for the case of a single $\pi$ bond in a
host of normal bonds, as calculated by this numerical procedure.  The
results are also compared to the total energy obtained from a
ground-state configuration corresponding to a pair of bound fractional
vortices of charge $\pm q$, calculated numerically.  As shown in the
figures, the agreement is excellent, thereby indicating that the
ground-state energy is indeed well characterized by a bound pair of
fractional vortices.

\begin{figure}[tb]
\epsfysize=8.1cm
\centerline{\epsffile{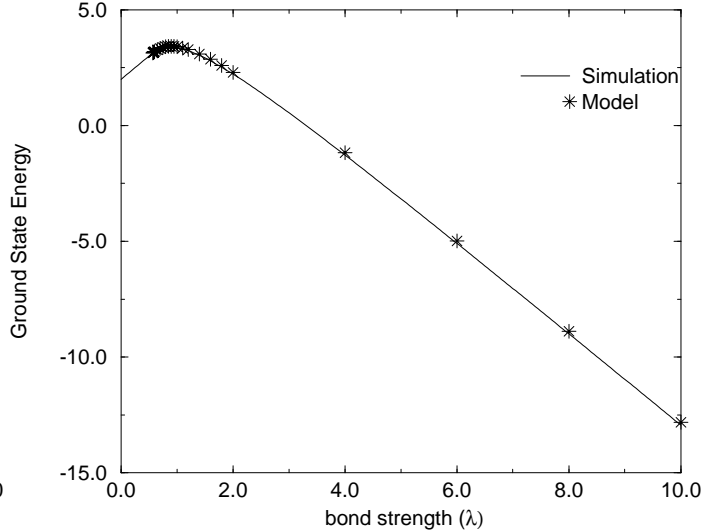}}
\caption{Same as Fig.\ \ref{fig10} but for two adjacent $\pi$ bonds.}
\label{fig12}
\end{figure}

Figs.\ \ref{fig11} and \ref{fig12} show a similar comparison for the
case of two $\pi$ bonds. Once again, the results obtained numerically
from the RSJ equations for both the total energy 

\noindent
and the bond angle across the $\pi$ bonds, are in excellent agreement
with those found from the fractional vortex ansatz, suggesting that
the ground state, in the case of two $\pi$ bonds, is again well
characterized, over a range of $\lambda$, by a pair of oppositely
charged fractional vortices.

\begin{figure}[tb]
\epsfysize=7cm
\centerline{\epsffile{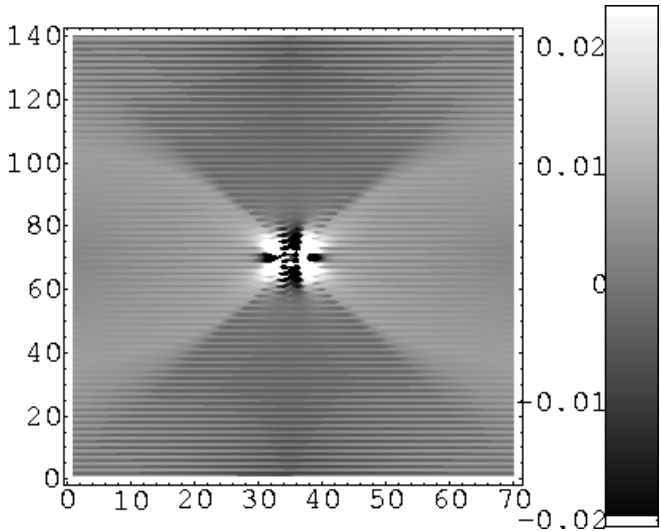}}
\vspace*{.2cm}
\caption{Graphical representation of the difference between the
      variational phase configuration (two oppositely charged
      fractional vortices) and the numerically obtained ground state
      phase configuration for the case: $\lambda = 10$, $N_\pi = 2$
      (string of two adjacent parallel $\pi$ bonds). The phase
      difference, as a fraction of the bond angle across the
      $\pi$ junctions, is shown on a gray scale given at the right
      edge of the diagram. A 70x70 lattice is considered and alternate
      stripes along the y-axis represent vertical and horizontal
      bonds. The $\pi$ bond string is shown in center of the Figure.}
\label{fig13}
\end{figure}

Finally, we briefly discuss the accuracy of our variational
approximation for the ``wave function,'' i.\ e., the phase
distribution in the ground state.  As is well known, a variational
wave function may give an excellent value for the ground state energy,
but a less accurate picture of the ground state configuration. In
particular, our variational approach ignores the spin-wave degrees of
freedom in characterising the ground state of the frustrated XY model,
but it is possible that they may be required to get an accurate
description of the phase distribution.  To test the accuracy of our
variational phase distribution, we have compared it to the exact
(numerical) phase distribution in the ground state in several cases.
The difference between the two configurations is shown graphically in
Fig.\ \ref{fig13} for the case of two $\pi$ bonds.  As can be seen,
the difference between the variational and exact wave functions is
almost always less than 2-3\% of the bond angle at the $\pi$ junction.
We have looked at the results for one and two $\pi$ bonds for varying
bond strengths, and in all cases considered the discrepancy is small.
Thus, we conclude that the phase distribution as well as the energy is
well approximated by our variational ground state involving only
fractional vortices.

\section{Summary and Possible Significance}

The original motivation for this work was to study $\pi$ bonds in
relation to the experiments of Kirtley {\it et al}
\cite{kirtley1,kirtley2,tsuei} on $\pi$-{\em grain boundaries} in
high-T$_c$ superconductors.

\noindent
Sigrist and Rice\cite{sigrist2} have shown
that the Josephson coupling across a grain boundary between two
$d$-wave superconductors can have either sign, depending on their
crystallographic orientations, thereby giving rise to the possibility
of $\pi$-grain boundaries. Of course, in the present model, we are
treating not a $\pi$-grain boundary, but rather a string of
$\pi$ bonds in the discrete XY model. Nevertheless, we argue that this
string could be viewed as a crude model of such a grain boundary.  It
has been argued that even so-called `single-crystal' high-T$_c$
superconductors can be effectively represented as an array of
superconducting grains weakly interacting via the Josephson coupling
between them \cite{tinkham}.  The typical lattice spacing for the
high-T$_c$ materials in such a model has been quoted to be as large as
1 $\mu m$.  Thus, the chain of $\pi$ bonds in our model can be taken
as representing the coupling of grains across a $\pi$-grain boundary,
and the length of the chain will depend on the dimensions of the grain
boundary, and the interpretation of the effective lattice spacing.

Now we turn to a summary of the experiments. The relevant experiments
fall into two categories. In the tricrystal experiments, the
intersections of three grain boundaries at a ``tricrystal point'' were
studied\cite{kirtley2,tsuei}.  At special orientations of the grain
boundaries, these experiments found that a half quantum of flux is
trapped around the tricrystal point - a result that has been
interpreted as verifying the $d$-wave symmetry of the superconducting
order parameter. In the tricrystal geometry, observing a trapped
half-flux quantum can then be explained by the fact that one of the
grain boundaries can be taken to be a $\pi$-boundary
\cite{bailey,geshkenbein}. In the second class of experiments, a
triangular (or a hexagonal) single-crystal superconductor was inserted
into a single crystal superconducting host of the same material, but
with crystal axes misoriented with respect to those of the inclusion.
In these systems, Kirtley {\it et al}\cite{kirtley1} have found
evidence of {\em fractional} (not half-integer) flux entrapment.
These results have been interpreted\cite{sigrist4} as evidence that
the superconducting order parameter violates time-reversal symmetry,
either in the bulk or at an interface.  Indeed, recent experiments
have reported fractional flux entrapment even in the absence of
$\pi$-grain boundaries \cite{tafuri}, possibly supporting the
existence of an order parameter which violates time-reversal symmetry.

If, in the triangular inclusion, only one of the three boundaries is a
$\pi$-boundary, the two ``zero'' boundaries will have little effect on
the arrangement of the order parameter phases, and can reasonably be
ignored.  Similarly, in the tricrystal, if only one of the three grain
boundaries is a $\pi$-boundary, this boundary would correspond to a
semi-infinite chain of $\pi$ bonds, while the other two ``zero''
boundaries can again be ignored in the model.  Thus, a finite chain of
$\pi$ bonds may be suitable for modeling the triangular inclusions,
and the extrapolation for long chain lengths is relevant for the
tricrystal experiments.

Next, we speculate about the relationship of our results to the
observed trapping of non-half-integers of flux in triangular
inclusions.  The trapped flux is usually related to the phase
difference across the grain boundary by the following
argument\cite{sigrist1}, which we restate to apply to our geometry.
Consider a closed integration contour $C$ (of radius $r\gg a$)
centered at one end of the grain boundary, and passing through the
grain boundary.  We wish to consider the flux enclosed by this loop.
The path is taken to be deep inside the grains, so that the Meissner
effect dictates that the supercurrent density ${\bf j} = 0$.  Since
${\bf j} \propto {\bf \nabla}\phi -(2\pi/\Phi_0){\bf A}$, where $\phi$
is the phase of the superconducting wave function, $\Phi_0=h c/ (2 e)$
is the superconducting flux quantum, and ${\bf A}$ is the vector
potential, it follows that
\begin{equation}
\label{eq:meissner}
{\bf \nabla} \phi = \frac{2 \pi}{\Phi_0} {\bf A}.
\end{equation} 
Now let $C = C^1 + C^2$, where $C^1$ is the part of the contour not
including the grain boundary.  In the approximation that $C^2$ can be
taken to be infinitesimally short, the integral $\int_{C^2}{\bf
A}\cdot{\bf d\ell} \sim 0$.  In addition, we have $\int_{C^2}{\bf
\nabla}\phi\cdot{\bf d\ell} = \Delta\phi$, the phase discontinuity
across the grain boundary.  But also $\phi$ must be continuous around
$C$, modulo $2\pi$.  Combining all these conditions with Eq.\
(\ref{eq:meissner}), we find that
\begin{equation}
\label{eq:fractional}
\Delta \phi = 2 \pi n - \frac{2 \pi}{\Phi_0} \Phi ,
\end{equation}
where $n$ is an integer and $\Phi$ is the flux enclosed by the {\em
entire} contour.  Thus, the flux enclosed by $C$ is related 
\newpage

\widetext
\begin{figure}[tb]
\epsfysize=7.5cm
\centerline{\epsffile{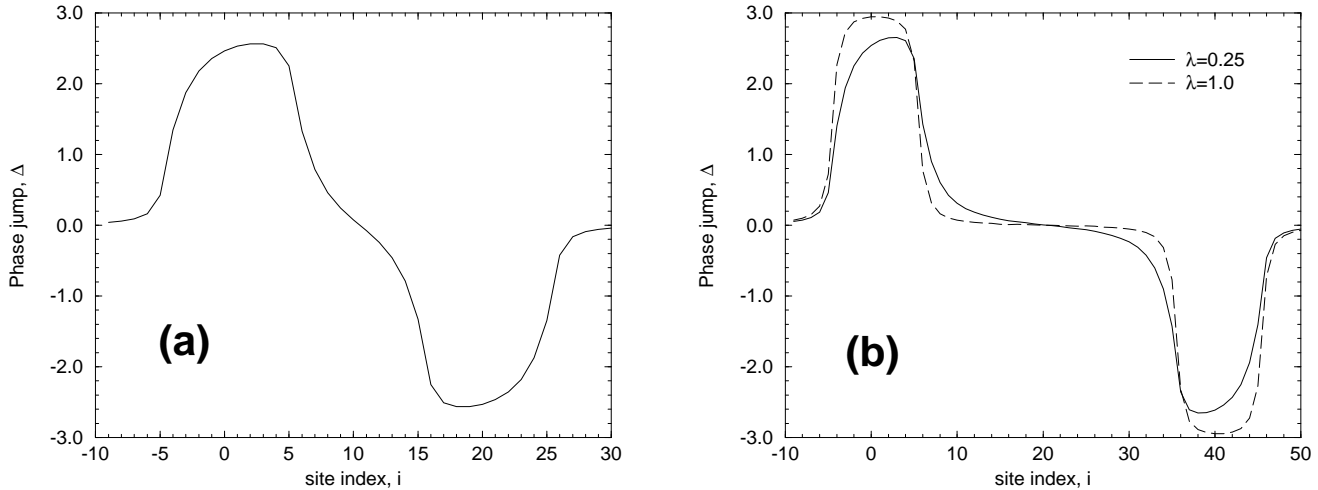}}
\caption{The calculated phase jump, $\Delta$, which corresponds to
      the flux measured by a SQUID of size 10 $\times$ 10 lattice
      spacings, for a chain of $\pi$ bonds of length $n$ and bond
      strength $\lambda$.  The x coordinate denotes the position of
      the center of the SQUID relative to the leftmost $\pi$ bond.
      $\Delta$ is defined to be the sum, taken in the counterclockwise
      direction, of the phase discontinuities across those bonds
      intersected by the perimeter of the SQUID, which are either
      $\pi$ bonds in the grain boundary, or lie along the extension of
      the grain boundary along the $\pm x$ direction.  (a) $n = 20$,
      $\lambda = 0.25$.  (b) $n = 40$, $\lambda = 0.25$ (full curve)
      and $1.0$ (dashed curve).}
\label{fig14}
\end{figure}
\narrowtext

\noindent
to the {\em phase defect} across the $\pi$ junction in the loop. In
particular, if $\Delta \phi$ is a {\em non-half-integer fraction} of
$2 \pi$ in the ground state, then the flux enclosed will also
correspondingly be fractional.  Hence, a non-half-integer fractional
flux is correlated with a phase jump across the grain boundary which
is a non-integer fraction of $\pi$.

Our results show that $\Delta\phi \neq \pi$ for an interior bond in a
finite chain of $\pi$ bonds. In fact $\Delta\phi$ can be `tuned' to be
any fraction of $\pi$ by simply varying the strength of the bonds for
any finite chain length. Thus, a necessary condition for the
occurrence of a non-half-integer flux quantum is indeed satisfied.
But this result still does not demonstrate that the the trapped
vortices correspond to non-half-integer flux quanta, because our
calculations do not include the magnetic fields induced by the
currents near the $\pi$-grain boundary.  These fields will change
${\bf A}$, and hence, the phase arrangement itself to some extent.
Thus, we cannot rigorously infer the magnetic flux when these
inductive effects are omitted from the calculations \cite{jose}.

Although our present calculations do not include these inductive
effects, it is still instructive to look at the phase distribution as
if Eq.\ (\ref{eq:fractional}) were valid anyway.  In particular, let
us try to model the flux configuration obtained by scanning the
$\pi$-grain boundary using an idealized square SQUID.  We take the
flux through the SQUID to be the same as that through the
corresponding contour $C$ as described above.  According to the
argument just given, the flux passing through the SQUID is therefore
proportional to the sum of the phase jumps $\Delta = \sum_i
\Delta\phi_i$ around the SQUID contour, across those bonds for which
the phase has a discontinuity.  In our simplified model for the flux
\newline
\newline

\vspace*{9.35cm}

\noindent
through the SQUID, these discontinuities occur across the two bonds
where the contour, taken counterclockwise around the SQUID, intersects
the $\pi$-grain boundary or its extension along the x axis.

In order to make a reasonable connection to the experimental geometry,
we estimate the lattice spacing $a$ in our model using
\begin{equation}
E_J = I_c \Phi_0 /c = a^2 J_c \Phi_0/c,
\end{equation}
where $E_J$ is the Josephson coupling energy between adjacent grains,
$I_c$ is the associated intergranular critical current, $J_c$ is the
macroscopic critical current density set by the Josephson effect
coupling, and $a$ is the lattice spacing of the granular array.

Using the experimental estimates for $E_J$ and $J_c$ (see Ref.\ $13$),
we estimate $a = 1.1 \mu m$, which is in agreement with the typical
value for these materials \cite{tinkham}. Since the triangular
insertions are roughly $ 20 \mu m$ in length (for each side), we have
looked at the results for a $\pi$ bond chain of length 20 lattice
spacings, with the SQUID diameter also taken to correspond to the
experimental diameter. In Fig.\ \ref{fig14}(a), we show $\Delta$ as
calculated for a $\pi$ bond chain of length $n=20$ and strength
$\lambda = 0.25$, and a SQUID of diameter $d=10$.  In Fig.\
\ref{fig14}(b), we show the same for a chain of length $n=40$. We note
that for a fixed bond strength, $\Delta$ becomes more concentrated
near the chain ends with increasing chain length $n$. Furthermore, we
can also make $\Delta$ more localized near the chain ends by
increasing $\lambda$, as shown in Fig.\ \ref{fig14}(b).

The profile of $\Delta$, shown in Fig.\ \ref{fig14}(a), strikingly
resembles the {\em flux profile} measured in Ref.\ $13$
across one side of a triangular insertion. In view of this similarity,
it would be interesting to see if the changes we see with bond
strength and chain length are also found experimentally.
(Experimentally, the strength of the bonds can be changed by varying
the misorientation angles of the inclusions.)  If inductive effects do
not change the qualitative picture presented above, then these
experiments would provide evidence supporting the interpretation of
the grain boundary as a string of $\pi$ bonds.

We also comment on the fact that Kirtley {\it et al}\cite{kirtley1}
were able to reproduce their measured flux configuration with a
certain arrangement of fractional magnetic charges.  Our picture
suggests one way of understanding {\em why} this modeling works.  The
key is that the flux distribution is closely related to the {\em
gauge-invariant} phase jump ($\Delta \gamma_{ij}$) across the
boundary, given by
\begin{equation}
\Delta \gamma_{ij} = \Delta \phi_{ij} -  (2\pi/\Phi_0)\Delta A_{ij},
\end{equation}    
where $i,j$ label sites connected by a bond across the grain boundary,
and $\Delta A_{ij}$ and $\Delta \phi_{ij}$ are the corresponding
discontinuities in the vector potential and the phase across that
bond. A nonzero $\Delta\gamma_{ij}$ can therefore be attributed either
to a nonzero $\Delta\phi_{ij}$ or a nonzero $\Delta A_{ij}$ (or a
combination of both).  In our calculations, we have assumed a nonzero
$\Delta\phi_{ij}$ and take $\Delta A_{ij} = 0$ across the branch cut.
The opposite choice ($\Delta A_{ij} \neq 0$, $\Delta \phi_{ij} = 0$)
corresponds to a fractional magnetic monopole, since the vector
potential of a magnetic monopole changes discontinuously across a
branch cut\cite{wilczek}.  Since the physical quantity is the
gauge-invariant phase difference, these pictures are equivalent.

In summary, we have introduced a set of ``fractional vortex ''
excitations in the XY model, and have derived an expression for the
interaction energy of a bound pair of fractional vortices.
Furthermore, we have studied the ground state of the XY model on a
two-dimensional lattice containing $\pi$ bonds. For strings of
$\pi$ bonds of any length, we find that there exists a minimum bond
strength, above which the ground state can be characterized by pair(s)
of oppositely charged fractional vortices. We have verified this
ansatz by carrying out independent numerical simulations for the
ground-state configuration of this system. Finally, we have discussed
the possible connection between these calculations and the trapped
fractional flux quanta, which are observed near grain boundaries in
high-T$_c$ superconductors.

\section{Acknowledgments}
We would like to thank Kathryn Moler, Yong-Baek Kim and Rajiv Singh
for helpful conversations.  We are grateful for support from NSF grant
DMR97-31511, the Midwest Superconductivity Consortium through Purdue
University. Grant No. DE-FG 02-90 45427, and NASA, Division of
Microgravity Sciences, Grant No. NCC 8-152.  RVK would also like to
acknowledge the support of the US Department of Energy, Office of
Science, Division of Materials Research. Calculations were carried out
using the IBM SP2 and the ORIGIN 2000 at the Ohio Supercomputer
Center.

\end{document}